\documentclass{ws-mpla}

\def\be{\begin{equation}}
\def\ee{\end{equation}}
\def\ba{\begin{eqnarray}}
\def\ea{\end{eqnarray}}
\begin{document}

\markboth{J. P. Beltr\'an Almeida, Y. Rodr\'{\i}guez, and C. A. Valenzuela-Toledo}
{The Suyama-Yamaguchi consistency relation in the presence of vector fields}

\catchline{}{}{}{}{}

\title{THE SUYAMA-YAMAGUCHI CONSISTENCY RELATION IN THE PRESENCE OF VECTOR FIELDS}

\author{\footnotesize JUAN P. BELTR\'AN ALMEIDA}

\address{Centro de Investigaciones, Universidad Antonio Nari\~no,  
\\ Cra 3 Este \# 47A-15, Bogot\'a D.C. 110231, Colombia\\
juanpbeltran@uan.edu.co}

\author{YEINZON RODR\'IGUEZ}

\address{Centro de Investigaciones, Universidad Antonio Nari\~no,  
\\ Cra 3 Este \# 47A-15, Bogot\'a D.C. 110231, Colombia,\\
and \\
Yukawa Institute for Theoretical Physics, Kyoto University,\\
 Kitashirakawa-Oiwake-cho, Sakyo-ku, Kyoto 606-8502, Japan,\\
and \\ 
Escuela de F\'isica, Universidad Industrial de Santander,  \\ 
Ciudad Universitaria, Bucaramanga 680002, Colombia\\
yeinzon.rodriguez@uan.edu.co
}

\author{C\'ESAR A. VALENZUELA-TOLEDO}

\address{Departamento de F\'isica, Universidad del Valle, \\
Ciudad Universitaria Mel\'endez, Santiago de Cali 760032, Colombia\\
cesar.valenzuela@correounivalle.edu.co
}
\maketitle

\pub{Received 22 Nov 2012}{Revised (Day Month Year)}

\begin{center}
PI/UAN-2011-512FT
\end{center}

\begin{abstract}
We consider inflationary models in which vector fields are responsible for part or eventually all of the primordial curvature perturbation $\zeta$.  Such models are phenomenologically interesting since they naturally introduce anisotropies in the probability distribution function of the primordial fluctuations that can leave a measurable imprint in the cosmic microwave background. Assuming that non-Gaussianity is generated due to the superhorizon evolution, we use the $\delta N$ formalism to do a complete tree level calculation of the 
non-Gaussianity parameters $f_{\rm NL}$ and $\tau_{\rm NL}$ in the presence of vector fields. 
We isolate the isotropic pieces of the non-Gaussianity parameters, 
which anyway have contributions from the vector fields, and show that they obey the Suyama-Yamaguchi consistency relation $\tau^{\rm iso}_{\rm NL}\geqslant(\frac{6}{5}f^{\rm iso}_{\rm NL})^2$. Other ways of defining the non-Gaussianity parameters, which could be observationally relevant, are stated and the respective Suyama-Yamaguchi-like consistency relations are obtained. 

\end{abstract}

\ccode{PACS No.: 98.80.Cq}

\section{Introduction}
The study of non-Gaussianities in the primordial curvature perturbation $\zeta$ is a subject of major interest in modern cosmology because the evaluation of the non-Gaussianity (NG) parameters provide criteria to discriminate among the many models proposed to explain the origin of the large-scale structure that we observe today\cite{obs,obs1,obs2}. Particular attention has been paid to the study of a consistency relation between the $f_{\rm NL}$ and $\tau_{\rm NL}$ parameters\cite{cr10,cr11,cr12,cr13,cr14,cr20,cr21}, the Suyama-Yamaguchi (SY) consistency relation ($\tau_{\rm NL} \geqslant (\frac{6}{5} f_{\rm NL})^2$ at least at tree level), because its violation would rule out many of the popular models of inflation. So far, it has been shown that this consistency relation works, in principle, on  models that only include scalar fields\cite{cr10,cr11,cr12,cr13,cr14} under just a few assumptions, but there is no any conclusive result for models that include other type of fields, for instance, vector fields (VF)\footnote{VF naturally generate scale-dependence and, therefore, the proofs in Ref. \cite{cr20,cr21} do not apply to those scenarios where VF contribute to the generation of the primordial curvature perturbation.}.  Vector field models are particularly interesting because they are suitable candidates to explain the apparent violation of statistical isotropy observed in recent analysis of data from the WMAP satellite\cite{gr1,gr2}, given that VF define inherently a preferred direction for the expansion, for the distribution of primordial fluctuations, or for both\cite{vi1,vi2,vi3,vi4,vi5,vi6,ValenzuelaToledo:2011fj,dklr}. Because of this reason, it is pertinent to study consistency relations in models that include scalar fields as well as vector fields as these not only include NG parameters but also the amount of primordial statistical anisotropy $g_\zeta$ \cite{varios1,varios2,varios3,varios4}. 


The main purpose of this paper is to find out how the well known SY consistency relation, valid for models involving only scalar fields under just a few assumptions, is modified when including VF. To this end, we do a complete tree level calculation of the $f_{\rm NL}$ and $\tau_{\rm NL}$ parameters based on the $\delta N$ formalism\cite{dklr}, and concentrate on the isotropic pieces of these parameters since these are the ones that are currently constrained by observations.  Under almost the same few assumptions made in the scalar fields case, and although the VF do contribute to the isotropic pieces of the NG parameters, we prove that the SY consistency relation is still obeyed, at least at tree level. Finally, the modified SY consistency relations for other ways of defining the NG parameters, which could be observationally relevant when taking into account the level of statistical anisotropy, are obtained. 

\section{The Curvature Perturbation}
In the presence of VF, part or even all of the primordial curvature perturbation $\zeta$ can be generated by vector field perturbations.  In such scenarios, the curvature perturbation can be calculated through the $\delta N$ formalism as an expansion in the perturbations of the fields a few e-folds after horizon exit\cite{dklr}: 
\be
 \zeta  \equiv \delta N = N_{a} \delta A_a  + \frac{1}{2}N_{a b} \delta A_a \delta A_b +  \frac{1}{3!}N_{ abc} \delta A_a \delta A_b \delta A_c + \cdots\ ,
 \ee
where $N$ is the number of e-folds during inflation and $N_{ab\cdots c}$ are the derivatives of $N$ with respect to $A_a$. In the previous expression, and in order to avoid notational clustering, we have assumed that there is only one scalar and one vector field; thus, $\delta A_{a} \equiv \delta \phi$ for $a=0$, and $\delta A_{a}\equiv \delta A_{i}$, the latter being the spatial components of the vector field, for $a=i=1,2,3$ \cite{ValenzuelaToledo:2011fj}, where these field perturbations are calculated in a flat slicing (the threading must be the comoving one). The extension of the results found in this letter to the case of several scalar and vector fields is straightforward.

\section{Calculation of the Correlators}\label{correlators}
\subsection{ The power spectrum} 
The power spectrum (PS) of the primordial curvature perturbation $\zeta$ is defined as:
\ba\label{twopoint}
\langle \zeta({\vec k}_1) \zeta({\vec k}_2) \rangle &=&(2\pi)^3 \delta({\vec k}_{12}) P_{\zeta}({\vec k}_1)\\ 
&=&(2\pi)^3 \delta({\vec k}_{12}) \frac{2 \pi^2 }{k^3}{\cal P}_{\zeta} ({\vec k}_1) \,,
\ea
where ${\cal P}_{\zeta} ({\vec k})$ is the {\it dimensionless power spectrum}, and ${\vec k}_{12} = {\vec k}_{1} + {\vec k}_{2}$. The statistical anisotropy in $\zeta$, due to the existence of a preferred direction $\hat{n}$, can be parametrized by a function $g_\zeta$ as follows\cite{acw}: 
\be  { P}_{\zeta}(\vec{k})= { P}_{\zeta}
^{{\rm iso}}(k)\left(1+g_\zeta(\hat{k}\cdot \hat{n})^2 \right),
\ee
where ${ P}_{\zeta}^{{\rm iso}}(k)$ denotes the isotropic part of the PS and the vectors $\hat{k}$ and $\hat{n}$ are of unit norm. 

The first assumption we will make is that the NG in the field perturbations is negligible, i.e., the NG is generated due to the superhorizon evolution.
The second assumption is to consider that the field perturbations\footnote{We refer, in the case of VF, to the scalar perturbations that multiply the respective polarization vectors in a polarization mode expansion.} are statistically isotropic, which is equivalent to the fact that there is no correlation between scalar and vector fields and that the expansion is isotropic\cite{ValenzuelaToledo:2011fj}. This may be reasonable as a good approximation in several cases\cite{dklr}, although it must be proven in each particular case, and is essential in the development of this paper because, otherwise, there would not exist any isotropic contributions to the NG parameters which are what really is constrained by observations. We shall adopt the notation
\be
 \langle \delta A_a (\vec {k}_1)  \delta A_b(\vec {k}_2)  \rangle = (2\pi)^3 \delta({\vec k}_{12}) \Pi_{ab}(\vec{k}_1) \,,
\ee
with the components
\ba \langle \delta A_0 (\vec {k}_1)  \delta A_0(\vec {k}_2)  \rangle &=& (2\pi)^3 \delta({\vec k}_{12}) P_{\delta \phi}({k}_1) \,, \\ 
\langle \delta A_i (\vec {k}_1)  \delta A_j(\vec {k}_2)  \rangle &=& (2\pi)^3 \delta({\vec k}_{12}) \Pi_{ij}(\vec{k}_1) \,,
\ea
 and zero otherwise. In the latter expression:
\begin{equation}
\Pi_{ij}(\vec{k}) = \Pi^{\rm even}_{ij}(\vec{k}) P_{+}(k) + \Pi^{\rm odd}_{ij}(\vec{k}) P_{-}(k) + \Pi^{\rm long}_{ij}(\vec{k}) P_{{\rm long}}(k) \,,
\end{equation}
where $\Pi^{\rm even}_{ij}(\vec{k}) = \delta_{ij} - \hat{k}_{i}\hat{k}_{j}, \ \Pi^{\rm odd}_{ij}(\vec{k}) = i \epsilon_{ijk}\hat{k}_{k}$, and
$\Pi^{\rm long}_{ij}(\vec{k}) =  \hat{k}_{i}\hat{k}_{j}$,  $P_{\rm long}(k)$ being the longitudinal power spectrum and $P_{+}(k)$ and $P_{-}(k)$ being the parity conserving and violating power spectra respectively\cite{ValenzuelaToledo:2011fj,dklr}. The fact that none of $P_{\delta \phi}(k)$, $P_{+}(k)$, $P_{-}(k)$, and $P_{\rm long}(k)$ depend on the direction of the wavevector is a consequence of the second assumption.  

At tree level, the PS is\cite{ValenzuelaToledo:2011fj,dklr}:
\be \label{ps-zeta}
P_{\zeta}(\vec{k}) = N_{a} N_{b} \Pi_{ab}(\vec{k}) \,.
\ee
For concreteness, we shall restrict to the case of vector field perturbations which preserve parity, then, we  just keep the $P_{+}(k)$ and  $P_{{\rm long}}(k)$ parts of the vector field power spectra. We also suppose that all the $\delta A_{i}$ perturbations evolve in the same way after horizon crossing so that they have the same spectral index in their power spectra. In  such conditions, the longitudinal and the parity conserving power spectra are related by $P_{{\rm long}}(k) = r P_{+}(k)$ where $r$ is a scale independent number. With these considerations in mind, the vector power spectra are:
\be \label{vps}
\Pi_{ij}(\vec{k}) = \left[ \delta_{ij} + (r-1) \hat{k}_{i}\hat{k}_{j}\right] P_{+}(k) \,.
\ee
Our next steps are to evaluate the PS by employing the $\delta N$ expansion and to separate the isotropic and anisotropic pieces obtaining in this way:
\ba
\label{ps-zeta-tilde}
{ P}_{\zeta}(\vec{k})= {N}_{a}  {N}_{b} { P}_{ab}(k)  + (r-1)\left({N}_{i} \hat{k}_{i} \right)^2 { P}_{+}(k) = { P}^{\rm iso}_{\zeta}(k)\left(1  + g_\zeta(\hat{n}_{i} \hat{k}_{i})^2 \right) \,,
\ea
where 
\begin{eqnarray}
{ P}^{\rm iso}_{\zeta} (k) &=& {N}_{a} {N}_{b} { P}_{ab}(k) \,, \label{isops} \\
{ P}^{\rm aniso}_{\zeta} (k) &=& (r-1)\left({N}_{i} \hat{k}_{i} \right)^2 { P}_{+}(k) \,, \label{anisops}
\end{eqnarray}
$\hat{n}_i = N_i/(N_j N_j)^{1/2}$ and the relation between the $(r-1)$ factor and the anisotropy parameter $g_\zeta$ in the power spectrum is given by:
\be\label{g-anis}
g_\zeta= (r-1)\frac{{N_i N_i} {\cal P}_{+} }{{N}_{a}  {N}_{b} {\cal P}_{ab}}.
\ee
In the latter expression, the dimensionless power spectra $ {\cal P}_{ab}$ assume the values $ {\cal P}_{00} =  {\cal P}_{\delta \phi}$ and  ${\cal P}_{ij} =  \delta_{ij}{\cal P}_{+}$.

\subsection{The bispectrum.} 
The bispectrum (BS) of the primordial curvature perturbation is defined as:
\ba
\langle \prod_{i=1}^3\zeta({\vec k}_i)\rangle &=& (2\pi)^3 \delta ({\vec k}_{123}) B_{\zeta}({\vec k}_1 ,\,{\vec k}_2 ,\, {\vec k}_3 )\\ 
 &=& (2\pi)^3  \delta ({\vec k}_{123}) \frac{4 \pi^4 }{k_1^3 \, k_2^3}{\cal B}_{\zeta} ({\vec k}_1 ,\,{\vec k}_2 ,\, {\vec k}_3) \,.
\ea
Using the $\delta N$ expansion we get the tree level BS:
\ba \label{bs-zeta}
B_{\zeta}=  N_{a} N_{b} N_{cd}\left[ \Pi_{ac}(\vec{k}_1) \Pi_{bd}(\vec{k}_2) + {\mbox 2\, {\rm perm.}} \right] \,.
\ea
Expanding the above by employing the expression in (\ref{vps}) and separating the tree level BS into
the isotropic and the anisotropic pieces we get
\ba \label{bs-1loop-iso}
{{ B}}_{\zeta}^{\rm iso}({ k}_1 ,\,{ k}_2 ,\, { k}_3 ) &=&    N_{a} N_{b} N_{cd}{\cal P}_{ac}{\cal P}_{bd}  \sum_{l<m} (k_l k_m)^{-3} \,,\\
\label{bs-1loop-aniso} \nonumber
{{ B}}_{\zeta}^{\rm aniso}({\vec k}_1 ,\,{\vec k}_2 ,\, {\vec k}_3 ) &=&    (r-1)N_{a} N_{i} N_{bj}{\cal P}_{ab}{\cal P}_{+} \sum_{l<m}\, 
(k_l k_m)^{-3}(\hat k_{(l)i} \hat k_{(l)j} + \hat k_{(m)i} \hat k_{(m)j} ) \\
&+& (r-1)^2N_{i} N_{k} N_{jn}{\cal P}^2_{+} \sum_{l<m}(k_l k_m)^{-3} \hat k_{(l)i} \hat k_{(l)j}\hat k_{(m)k} \hat k_{(m)n} \,.
\ea

\subsection{The trispectrum} 
Finally, we evaluate the trispectrum (TS) of the primordial curvature perturbation:
\ba
&&\langle \prod_{i=1}^4\zeta({\vec k}_i)\rangle=(2\pi)^3 \delta ({\vec k}_{1234} ) T_{\zeta}({\vec k}_1 ,\,{\vec k}_2 ,\, {\vec k}_3 ,\, {\vec k}_4 ) \\ 
&&=(2\pi)^3 \delta ({\vec k}_{1234} ) \frac{8 \pi^6 }{k_1^3 \, k_2^3 \, k_{23}^3}{\cal T}_{\zeta} ({\vec k}_1 ,\,{\vec k}_2 ,\, {\vec k}_3 ,\, {\vec k}_4) \,. \nonumber \\
&&
\ea
Once again, we use the $\delta N$ expansion to get the tree level contribution to the TS, which results in
\ba 
&&T_{\zeta} = T^{\tau_{\rm NL}}_{\zeta} + T^{g_{\rm NL}}_{\zeta}\\ \nonumber
&&=  N_{a} N_{b} N_{cd} N_{ef} \left[\Pi_{ac}(\vec{k}_1) \Pi_{be}(\vec{k}_2)\Pi_{df}(\vec{k}_{13}) + {\mbox 11\, {\rm perm.}} \right] \\ 
&&+ N_{a} N_{b} N_{c} N_{def}  \left[\Pi_{ad}(\vec{k}_1) \Pi_{be}(\vec{k}_2)\Pi_{cf}(\vec{k}_3) + {\mbox 3\, {\rm perm.}} \right] \,. \nonumber \\
&&
\ea
As we are interested in the $\tau_{\rm NL}$ amplitude, we only need the term $T^{\tau_{\rm NL}}_{\zeta}$ given that the term $T^{g_{\rm NL}}_{\zeta}$  enters in the definition of the $g_{\rm NL}$ parameter. The decomposition of $T^{\tau_{\rm NL}}_{\zeta}$ in isotropic and anisotropic pieces is given by
{\small\ba\label{ts-iso}
{ T}_{\zeta}^{\rm iso (\tau_{\rm NL})} &=&  N_{a} N_{b} N_{cd} N_{ef} {\cal P}_{ac} {\cal P}_{be}{\cal P}_{df} \,  \sum_{l<m,  s\neq m, l} (k_l k_m k_{ls})^{-3} \,,  \\ \label{ts-aniso} \nonumber
{ T}_{\zeta}^{\rm aniso (\tau_{\rm NL})} &=& (r-1) N_{i} N_{b} N_{jd} N_{ef} {\cal P}_{be} {\cal P}_{df}{\cal P}_{+}  \sum_{l<m, s\neq m, l} (k_l k_m k_{ls})^{-3}  (\hat k_{(l)i} \hat k_{(l)j} + \hat k_{(m)i} \hat k_{(m)j})  \\ \nonumber
& + & (r-1) N_{b} N_{d} N_{ei} N_{fj} {\cal P}_{be} {\cal P}_{df}{\cal P}_{+}  \sum_{l<m, s\neq m, l} (k_l k_m k_{ls})^{-3}  \hat k_{(ls)i} \hat k_{(ls)j}  \\ \nonumber
& + & (r-1)^2 N_{i} N_{k} N_{jd} N_{lf} {\cal P}_{df}{\cal P}^2_{+} \sum_{l<m, s\neq m, l} (k_l k_m k_{ls})^{-3}  \hat k_{(l)i} \hat k_{(l)j}\hat k_{(m)k} \hat k_{(m)n}  \\ \nonumber
& + & (r-1)^2 N_{i} N_{d} N_{jk} N_{fn} {\cal P}_{df}{\cal P}^2_{+} \sum_{l<m, s\neq m, l} (k_l k_m k_{ls})^{-3}  \hat k_{(ls)k} \hat k_{(ls)n}( \hat k_{(m)i} \hat k_{(m)j} + \hat k_{(l)i} \hat k_{(l)j})  \\  
& + & (r-1)^3 N_{i} N_{k} N_{jm} N_{hn} {\cal P}^3_{+}\sum_{l<m, s\neq m, l} (k_l k_m k_{ls})^{-3} \hat k_{(l)i} \hat k_{(l)j}\hat k_{(m)k} \hat k_{(m)h} \hat k_{(ls)m} \hat k_{(ls)n}. 
\ea}
\section{The NG Parameters $f_{\rm NL}$ and $\tau_{\rm NL}$}
In this work we consider the definition of the non-gaussianity parameters $f_{\rm NL}$ and $\tau_{\rm NL}$ as follows
\ba\label{f-totalPS}
B_{\zeta} &=& \frac{6}{5}f_{\rm NL} (P_{\zeta}({\vec k_1})P_{\zeta}({\vec k_2})+{\mbox 2\, {\rm perm.}}) \,, \\ \label{tau-totalPS}
T_{\zeta} &=& \tau_{\rm NL} (P_{\zeta}({\vec k_1})P_{\zeta}({\vec k_2})P_{\zeta}({\vec k_{14}})+{\mbox 11 \, {\rm perm.}}) \,.
\ea
Notice that, in the definitions above, we have used the full PS including both scalar and vector field perturbations which, accordingly, include isotropic and anisotropic contributions. Nevertheless, it is usual to find in the literature a definition which considers only the isotropic piece of the PS\cite{obs,obs1,obs2}, so the NG parameters are given by $B_{\zeta} = \frac{6}{5}f'_{\rm NL} (P_{\zeta}^{\rm iso}({\vec k_1})P_{\zeta}^{\rm iso}({\vec k_2})+{\mbox 2\, {\rm perm.}})$ and $T_{\zeta} = \tau'_{\rm NL} (P_{\zeta}^{\rm iso}({\vec k_1})P_{\zeta}^{\rm iso}({\vec k_2})P_{\zeta}^{\rm iso}({\vec k_{14}})+{\mbox 11 \, {\rm perm.}})$.
We can relate both definitions using the expression in (\ref{ps-zeta-tilde}) for the PS:
\ba\label{f-f'}
f_{\rm NL}&=& \frac{1}{1+\chi_1}f'_{\rm NL} = \frac{1}{1+\sum_{i=1}^2 m_i (g_{\zeta})^i}f'_{\rm NL} \,,  \\\label{tau-tau'}
\tau_{\rm NL}&=& \frac{1}{1+\chi_2}\tau'_{\rm NL} = \frac{1}{1+\sum_{i=1}^3 l_i (g_{\zeta})^i}\tau'_{\rm NL} \,,
\ea
where the parameters $m_i$ and $l_i$ are obtained from the expansion of the products of the PS in (\ref{ps-zeta-tilde}) which are in the definitions of the NG parameters. They depend on the derivatives of $N$ and the momenta $\vec{k}_i$. Explicitly:
{\small\ba\label{psps} 
&& P_{\zeta}({\vec k_1})P_{\zeta}({\vec k_2})+{\mbox 2\, {\rm perm.}} = (1+\chi_1) \left[P^{{\rm iso}}_{\zeta}({ k_1})P^{{\rm iso}}_{\zeta}({ k_2})+{\mbox 2\, {\rm perm.}}\right] \nonumber  \\  \nonumber
&& = \,\left[ 1+g_{\zeta}\,\hat{n}_i \hat{n}_j\,\frac{\sum_{l<m}\, 
(k_l k_m)^{-3} (\hat k_{(l)i} \hat k_{(l)j} + \hat k_{(m)i} \hat k_{(m)j} )}{ \sum_{l<m} (k_l k_m)^{-3}} \, \right. \\  
&& +   \left. g_{\zeta}^2\,\hat{n}_i \hat{n}_j\hat{n}_k \hat{n}_n\, \frac{\sum_{l<m}
 (k_l k_m)^{-3} \hat k_{(l)i} \hat k_{(l)j}\hat k_{(m)k} \hat k_{(m)n} }{\sum_{l<m} (k_l k_m)^{-3} } \right] (N_a N_b {\cal P}_{ab})^2  \sum_{l<m} (k_l k_m)^{-3} \,, \nonumber \\
&& 
\ea
\ba\label{pspsps} 
&& P_{\zeta}({\vec k_1})P_{\zeta}({\vec k_2})P_{\zeta}({\vec k_{13}})+{\mbox 11\, {\rm perm.}} = (1+\chi_2) \left[P^{{\rm iso}}_{\zeta}({ k_1})P^{{\rm iso}}_{\zeta}({ k_2})P^{{\rm iso}}_{\zeta}({ k_{13}})+{\mbox 11\, {\rm perm.}}\right]  \nonumber \\  \nonumber
&& = \,\left[ 1+g_{\zeta}\,\hat{n}_i \hat{n}_j\,\frac{\sum_{l<m,  s\neq m, l}\, 
(k_l k_m k_{ls})^{-3} (\hat k_{(l)i} \hat k_{(l)j} + \hat k_{(m)i} \hat k_{(m)j} + \hat k_{(ls)i} \hat k_{(ls)j} )}{ \sum_{l<m,  s\neq m, l} (k_l k_m k_{ls})^{-3}} \, \right. \\  \nonumber 
&& +   \left. g_{\zeta}^2\,\hat{n}_i \hat{n}_j\hat{n}_k \hat{n}_n\, \frac{\sum_{l<m,  s\neq m, l} 
 (k_l k_m k_{ls})^{-3} \left[\hat k_{(l)i} \hat k_{(l)j}\hat k_{(m)k} \hat k_{(m)n} + \hat k_{(ls)i} \hat k_{(ls)j}( \hat k_{(m)k} \hat k_{(m)n} + \hat k_{(l)k} \hat k_{(l)n})\right] }{\sum_{l<m,  s\neq m, l} (k_l k_m k_{ls})^{-3} } \, \right. \\ \nonumber 
&& +  \left. g_{\zeta}^3\,\hat{n}_i \hat{n}_j\hat{n}_k \hat{n}_h \hat{n}_t \hat{n}_n \, \frac{\sum_{l<m,  s\neq m, l}
(k_l k_m k_{ls})^{-3} \hat k_{(l)i} \hat k_{(l)j}\hat k_{(m)k} \hat k_{(m)h} \hat k_{(ls)t} \hat k_{(ls)n}  }{\sum_{l<m,  s\neq m, l} (k_l k_m k_{ls})^{-3} } \right] \nonumber \\
&& \times (N_a N_b {\cal P}_{ab})^3  \sum_{l<m,  s\neq m, l} (k_l k_m k_{ls})^{-3} \,.
\ea}
Certainly, both definitions of the NG parameters are approximately equal if the corrections arising from the existence of some statistical anisotropies were negligible.  Otherwise, such corrections would be very important from the observational point of view. 

\subsection{The $f_{\rm NL}$ parameter}
For the $f_{\rm NL}$ parameter we can use the results (\ref{isops}), (\ref{anisops}), (\ref{bs-1loop-iso}), and (\ref{bs-1loop-aniso}) to separate the isotropic and the anisotropic contributions as follows: $f_{\rm NL}=f_{\rm NL}^{\rm I}+f_{\rm NL}^{\rm II}={ B}^{\rm iso}_{\zeta}/({ P}_{\zeta}{ P}_{\zeta}+{\mbox 2 {\rm p.}})+{ B}^{\rm aniso}_{\zeta}/({ P}_{\zeta}{ P}_{\zeta}+{\mbox 2 {\rm p.}})$. The labels {\rm I} and
{\rm II} are related to the isotropic and the anisotropic pieces of the bispectrum in  (\ref{bs-1loop-iso}) and (\ref{bs-1loop-aniso}) respectively.
Using this decomposition, we can compare the $f_{\rm NL}^{\rm I}$ and $f_{\rm NL}^{\rm II}$ contributions by evaluating its ratio
{\ba \label{f-ratio}\nonumber
\frac{f_{\rm NL}^{\rm II}}{f_{\rm NL}^{\rm I}} &\equiv & \xi_1 \equiv \sum_{l=1}^2 f_l (g_{\zeta})^l\\ \nonumber
&=&g_{\zeta}\frac{ N_{a}N_{bj} N_{i}N_{c}N_{d}   {\cal P}_{ab}{\cal P}_{cd} }{N_{a} N_{b} N_{cd}N_{k}N_{k} {\cal P}_{ac}{\cal P}_{bd}}\frac{\sum_{l<m}\, 
(k_l k_m)^{-3}(\hat k_{(l)i} \hat k_{(l)j} + \hat k_{(m)i} \hat k_{(m)j} )}{ \sum_{l<m} (k_l k_m)^{-3}} \\
&+& g_{\zeta}^2\frac{ (N_{a}N_{b}{\cal P}_{ab})^2 N_{i} N_{k} N_{jn}}{(N_{i}N_{i})^2 N_{a} N_{b} N_{cd} {\cal P}_{ac}{\cal P}_{bd}} \frac{\sum_{l<m}
(k_l k_m)^{-3} \hat k_{(l)i} \hat k_{(l)j}\hat k_{(m)k} \hat k_{(m)n} }{\sum_{l<m} (k_l k_m)^{-3} }  \,.
\ea}  In the latter expression, we have used the expression in (\ref{g-anis}) to write the series expansion in terms of the level of statistical anisotropy $g_{\zeta}$. The coefficients $f_l$ can be read off directly from the latter expression.
Using the ratio $f_{\rm NL}^{\rm II}/f_{\rm NL}^{\rm I}$,  we can write the total $f_{\rm NL}$ as: 
$f_{\rm NL} = (1+\xi_1) f_{\rm NL}^{\rm I}=(1+\xi_1)/(1+\chi_1)f_{\rm NL}^{\rm iso}$, where $f_{\rm NL}^{\rm iso}$ is the piece  of 
$f_{\rm NL}$ that depends only on the isotropic BS in (\ref{bs-1loop-iso}) and the isotropic PS in (\ref{isops}). A very
appealing feature that we can see from the previous analysis is that the ratio of the terms related to the anisotropic and the isotropic parts of the NG parameter $f_{\rm NL}$
depends on the parameters of the model; for instance, it depends on the value of the ratio $r$, the level of statistical anisotropy $g_{\zeta}$, and the shape of the triangle defined by the $\vec{k}_i$ momenta. Moreover, interestingly enough, for some values of $g_{\zeta}$ and for some configurations
of the momenta,  the anisotropic contribution could be of the same order of magnitude as the isotropic piece and cannot be neglected. 

\subsection{The $\tau_{\rm NL}$ parameter.}
For the $\tau_{\rm NL}$ parameter, we apply the same decomposition related to the isotropic and anisotropic pieces that we did before with the $f_{\rm NL}$ parameter: $\tau_{\rm NL}= \tau^{\rm I}_{\rm NL}+\tau^{\rm II}_{\rm NL}={ T}^{\rm iso}_{\zeta}/({ P}_{\zeta}{ P}_{\zeta}{ P}_{\zeta}+{\mbox 11 {\rm p.}})+{ T}^{\rm aniso}_{\zeta}/({ P}_{\zeta}{ P}_{\zeta}{ P}_{\zeta}+{\mbox 11 {\rm p.}})$.  We follow the same steps that we did in the case of the $f_{\rm NL}$ parameter and write the ratio of both contributions as a series in powers of the $g_{\zeta}$
parameter:  %
{\small { \ba \nonumber
&& \frac{\tau_{\rm NL}^{\rm II}}{\tau_{\rm NL}^{\rm I}} \equiv   \xi_2 \equiv \sum_{l=1}^3 \tau_l (g_{\zeta})^l \\ \nonumber
&& = g_{\zeta}\frac{ (N_a N_b  {\cal P}_{ab}) N_{i} N_{c} N_{jd} N_{ef} {\cal P}_{ce} {\cal P}_{df}}{ (N_{i}N_{i}) N_{a} N_{b} N_{cd} N_{ef} {\cal P}_{ac} {\cal P}_{be}{\cal P}_{df}}  \frac{  \sum_{l<m, s\neq m, l} (k_l k_m k_{ls})^{-3}  (\hat k_{(l)i} \hat k_{(l)j} + \hat k_{(m)i} \hat k_{(m)j})   }{  \sum_{l<m,  s\neq m, l} (k_l k_m k_{ls})^{-3}   } \\ \nonumber
&& + g_{\zeta}\frac{ (N_a N_b  {\cal P}_{ab}) N_{b} N_{d} N_{ei} N_{fj} {\cal P}_{be} {\cal P}_{df}}{ (N_{i}N_{i}) N_{a} N_{b} N_{cd} N_{ef} {\cal P}_{ac} {\cal P}_{be}{\cal P}_{df}}  \frac{  \sum_{l<m, s\neq m, l} (k_l k_m k_{ls})^{-3}  \hat k_{(ls)i} \hat k_{(ls)j}   }{  \sum_{l<m,  s\neq m, l} (k_l k_m k_{ls})^{-3}   } \\ \nonumber
&&+ g_{\zeta}^2\frac{ (N_a N_b{\cal P}_{ab})^2 N_{i} N_{k} N_{jd} N_{nf} {\cal P}_{df}  }{ (N_{i} N_{i})^2 N_{a} N_{b} N_{cd} N_{ef} {\cal P}_{ac} {\cal P}_{be}{\cal P}_{df}} \frac{   \sum_{l<m, s\neq m, l} (k_l k_m k_{ls})^{-3}  \hat k_{(l)i} \hat k_{(l)j}\hat k_{(m)k} \hat k_{(m)n}   }{   \sum_{l<m,  s\neq m, l} (k_l k_m k_{ls})^{-3}   }\\ \nonumber
&&+ g_{\zeta}^2\frac{ (N_a N_b{\cal P}_{ab})^2 N_{i} N_{d} N_{jk} N_{fn} {\cal P}_{df}  }{ (N_{i} N_{i})^2 N_{a} N_{b} N_{cd} N_{ef} {\cal P}_{ac} {\cal P}_{be}{\cal P}_{df}} \frac{   \sum_{l<m, s\neq m, l} (k_l k_m k_{ls})^{-3}  \hat k_{(ls)k} \hat k_{(ls)n}( \hat k_{(m)i} \hat k_{(m)j} + \hat k_{(l)i} \hat k_{(l)j})  }{   \sum_{l<m,  s\neq m, l} (k_l k_m k_{ls})^{-3}   }\\ 
&&+ g_{\zeta}^3\frac{ (N_a N_b{\cal P}_{ab})^3 N_{i} N_{k} N_{jm} N_{hn}  }{(N_{i} N_{i})^3 N_{a} N_{b} N_{cd} N_{ef} {\cal P}_{ac} {\cal P}_{be}{\cal P}_{df}} \frac{  \sum_{l<m, s\neq m, l} (k_l k_m k_{ls})^{-3} \hat k_{(l)i} \hat k_{(l)j}\hat k_{(m)k} \hat k_{(m)h} \hat k_{(ls)m} \hat k_{(ls)n}   }{   \sum_{l<m,  s\neq m, l} (k_l k_m k_{ls})^{-3}   } \,.  \nonumber \\
&& \label{tau-ratio}
\ea} }
The coefficients $\tau_l$ can be read off directly from the expression in (\ref{tau-ratio}).
Then, the $\tau_{\rm NL}$ parameter can be written as:
$\tau_{\rm NL} = \left( 1 +\xi_2  \right) \tau_{\rm NL}^{\rm I}=(1+\xi_2)/(1+\chi_2)\tau_{\rm NL}^{\rm iso}$, where 
$\tau_{\rm NL}^{\rm iso}$ is the part of $\tau_{\rm NL}$ that depends only on the isotropic TS in (\ref{ts-iso}) and the isotropic PS in (\ref{isops}). The ratio again depends on the $g_{\zeta}$ parameter and on the configuration described by the $\vec{k}_i$ momenta.

\section{ A Consistency Relation for $f_{\rm NL}$ and $\tau_{\rm NL}$}
Searching for relations between the NG parameters, we start exploring their isotropic pieces. 
Here we give an explicit proof 
that the isotropic pieces of the NG parameters, in presence of scalar and vector field perturbations, obey the SY
consistency relation\cite{cr10,cr11,cr12,cr13,cr14}. Using the expressions in (\ref{isops}), (\ref{bs-1loop-iso}), and (\ref{ts-iso}) we find that:
\ba
\frac{6}{5}f_{\rm NL}^{\rm (iso)} &=& \frac{N_a N_b N_{cd}{\cal P}_{ac}{\cal P}_{bd}}{(N_a N_b {\cal P}_{ab})^2 } \,, \\
\tau_{\rm NL}^{\rm (iso)} &=& \frac{N_a N_b N_{cd}N_{ef}{\cal P}_{ac}{\cal P}_{be} {\cal P}_{df}}{(N_a N_b {\cal P}_{ab})^3 } \,.
\ea
Now, we can write the scale invariant power spectra ${\cal P}_{ab}$ as:
\begin{equation}
{\cal P}_{ab} = D_{ab}{\cal P}_{\delta\phi} = \left( \begin{array}{cc}
1 & 0 \\
0 & s \delta_{ij}
\end{array} \right)_{ab}{\cal P}_{\delta\phi} \,,
\end{equation}
where $s = {\cal P}_{+}/{\cal P}_{\delta\phi} $ is the vector to scalar PS ratio. Defining the vectors $K_a = D_{ab} N_b$ and $T_a = N_{ab} K_b$, we evaluate the ratio $A_{\rm NL}=\tau_{\rm NL}^{\rm iso}/(\frac{6}{5}f_{\rm NL}^{\rm iso})^2$:
\ba\label{cr-iso}
A_{\rm NL} =  \frac{(N_a K_a)(D_{ab} T_a T_b)}{(K_a T_a)^2} &=& \frac{(D_{ab}N_a N_b)(D_{ab} T_a T_b)}{(D_{ab} N_a T_b)^2} \geqslant 1\,, \nonumber \\
&&
\ea
which implies: 
\be
\tau^{\rm iso}_{\rm NL}\geqslant \left(\frac{6}{5}f^{\rm iso}_{\rm NL}\right)^2 \,.
\ee
The latter expression is a direct consequence of the Cauchy-Schwartz inequality for the vectors $N_a$ and $T_a$ with the positive 
definite metric $D_{ab}$, and reflects the fact that the SY consistency relation, at least at tree level, applies for the isotropic pieces of the NG parameters even in the presence of vector fields. 

Now, we can try to establish a relation for the complete NG parameters $f_{\rm NL}$ and $\tau_{\rm NL}$.  The fact that the isotropic NG parameters obey the SY inequality  implies that the total NG parameters obey a modified consistency relation. The modification to this relation will depend on several aspects, for instance, on the level of statistical anisotropy, the momenta configuration, etc. Employing the SY inequality for the isotropic parameters, we deduce that the total NG parameters obey the  relation:
\be\label{cr1}
\tau_{{\rm NL}} \geqslant \frac{(1+\xi_2)(1+\chi_1)^2}{(1+\chi_2)(1+\xi_1)^2}\left(\frac{6}{5}f_{{\rm NL}}\right)^2 \,.
\ee
On the other hand, the relation obeyed by the NG parameters $f'_{{\rm NL}}$ and  $\tau'_{{\rm NL}}$ is the following:
\be\label{cr2}
\tau'_{{\rm NL}} \geqslant \frac{(1+\xi_2)}{(1+\xi_1)^2}\left(\frac{6}{5}f'_{{\rm NL}}\right)^2 .
\ee
All of these relations represent a constraint on the NG and statistical anisotropy parameters for models of inflation that involve vector fields.

\section{Conclusions}

In this paper, we have studied under which conditions new kinds of the well known Sayama-Yamaguchi consistency relation are 
allowed  when vector fields are included in the inflationary dynamics. In particular we have shown that the isotropic pieces of the 
non-gaussianity parameters obey the Suyama-Yamaguchi consistency relation $\tau^{\rm iso}_{\rm NL}\geqslant(\frac{6}{5}f^{\rm 
iso}_{\rm NL})^2$ when its calculation is performed at tree level. 

We have also derived a modified consistency relation for the complete non-gaussianity parameters $f_{\rm NL}$ and $\tau_{\rm NL}$ and for a set of new parameters $f'_{\rm NL}$ and $\tau'_{\rm NL}$ that could be observationally relevant; these relations are given by the Eqs. \eqref{cr1} and \eqref{cr2}. Since the latter depend on the level of statistical anisotropy and the configuration of the wavevectors, the naive relations $\tau_{\rm NL}\geqslant(\frac{6}{5}f_{\rm NL})^2$ and $\tau'_{\rm NL}\geqslant(\frac{6}{5}f'_{\rm NL})^2$ could eventually be
violated in some particular inflationary model that includes vector fields. On this basis, we expect that in a near future the derived relations could provide a criteria to discriminate among the different models for the generation of the primordial curvature perturbation when vector fields are involved.

\section*{Acknowledgments}
J.P.B.A. and Y.R. are supported by Fundaci\'on para la Promoci\'on de la Investigaci\'on y la Tecnolog\'{\i}a del Banco de la Rep\'ublica (COLOMBIA) grant number 3025 CT-2012-02.  J.P.B.A. is supported by VCTI (UAN) grant number 2010251.  Y.R. was a JSPS postdoctoral fellow (P11323) and, in addition, is supported by DIEF de Ciencias (UIS) grant number 5177.  C.A.V.-T. is supported by Vicerrector\'{\i}a de Investigaciones (UNIVALLE) grant number 7858.


\end{document}